# Scalable multi-particle entanglement of trapped ions

H. Häffner[1,2], W. Hänsel[1], C. F. Roos[1,2], J. Benhelm[1,2], D. Chek–al–kar[1], M. Chwalla[1], T. Körber[1,2], U. D. Rapol[1,2], M. Riebe[1], P. O. Schmidt[1], C. Becher[1,†], O. Gühne[2], W. Dür[2,3] & R. Blatt[1,2]

[1] *Institut für Experimentalphysik, Universität Innsbruck, Technikerstraße 25, A–6020 Innsbruck, Austria*

[2] *Institut für Quantenoptik und Quanteninformation der Österreichischen Akademie der Wissenschaften, Technikerstraße 21a, A–6020 Innsbruck, Austria*

[3] *Institut für Theoretische Physik, Universität Innsbruck, Technikerstraße 25, A–6020 Innsbruck, Austria*

[†] *Present address: Fachrichtung Technische Physik, Universität des Saarlandes, Postfach 151150, D-66041 Saarbrücken, Germany*

**Entanglement, its generation, manipulation and fundamental understanding is at the very heart of quantum mechanics. The phrase *entanglement* was coined by Erwin Schrödinger in 1935 for particles that are described by a *common* wave function where individual particles are not independent of each other but where their quantum properties are inextricably interwoven[1]. Entanglement properties of two and three particles have been studied extensively and are very well understood. Entanglement of four[2] and five[3] particles was demonstrated experimentally. However, both creation and characterization of entanglement become exceedingly difficult for multi–particle systems. Thus the availability of such multi–particle entangled states together with the full information on these states in form of their**

**density matrices creates a test-bed for theoretical studies of multi-particle entanglement. Among the various kinds of entangled states, the W–state[4–6] plays an important role since its entanglement is maximally persistent and robust even under particle losses. Such states are central as a resource to the new fields of quantum information processing[7] and multi-party quantum communication. Here we report the deterministic generation of four-, five-, six-, seven- and eight–particle entangled states of the W–type with trapped ions. We obtain the maximum possible information on these states by performing full characterization via state tomography[8]. Moreover, we prove in a detailed analysis that they carry genuine four-, five-, six-, seven- and eight–particle entanglement, respectively.**

An $N$–particle W–state

$$|W_N\rangle = (|D\cdots DDS\rangle + |D\cdots DSD\rangle + |D\cdots DSDD\rangle + \cdots + |SD\cdots D\rangle)/\sqrt{N} \quad (1)$$

consists of a superposition of $N$ states where exactly one particle is in the $|S\rangle$–state while all other particles are in $|D\rangle$[4,5]. W–states are genuine $N$–particle entangled states of special interest: their entanglement is not only maximally persistent and robust under particle losses[9], but also immune against global dephasing, and rather robust against bit flip noise. In addition, for larger numbers of particles, W–states may lead to stronger non–classicality[10] than GHZ–states[11] and may be used for quantum communication[12,13].

The generation of such W–states is performed in an ion–trap quantum processor[14]. We trap strings of up to eight $^{40}$Ca$^+$ ions in a linear Paul trap. Superpositions of the S$_{1/2}$ ground state and the metastable D$_{5/2}$ state of the Ca$^+$ ions (lifetime of the $|D\rangle$–level: $\tau \approx 1.16$ s) represent the qubits. Each ion–qubit in the linear string is individually addressed by a series of tightly focused laser pulses on the $|S\rangle \equiv S_{1/2}(m_j = -1/2) \longleftrightarrow |D\rangle \equiv D_{5/2}(m_j = -1/2)$ quadrupole transition employing narrowband laser radiation near 729 nm. Doppler cooling and subsequent sideband cooling prepare the ion string in the ground state of the center–of–mass vibrational mode. Optical pumping initializes the ions' electronic qubit states in the $|S\rangle$ state. After preparing an entangled state with a series of laser pulses, the quantum state is read out with a CCD camera.

The W–states are efficiently generated by sharing one motional quantum between the ions with partial swap–operations (see Tab. 1)[6]. For an increasing ion number, however, the initialization of the quantum register becomes more and more difficult as imperfections —like optical pumping— add up for each ion. Therefore for $N = 6, 7, 8$, we first prepare the state $|0, DD \cdots D\rangle$ with $N$ carrier $\pi$ pulses, where the 0 refers to the motional state of the center–of–mass mode. Then, laser light resonant with the $S \leftrightarrow P$–transition projects the ion string on the measurement basis. Absence of fluorescence reveals whether all ions were prepared in $|D\rangle$. Similarly, we test the motional state with a single blue $\pi$ pulse. Absence of fluorescence during a subsequent detection period indicates ground state occupation. This initialization procedure can be viewed as a generalized optical pumping with the target state $|0, DD \cdots D\rangle$. If both checks were successful (total success rate $\geq 0.7$), we continue with the $|W\rangle$–preparation at step ($i3$) in Tab. 1. Thus, we create $|W_N\rangle$–states ($N \leq 8$) in about $500 - 1000$ $\mu$s.

Full information on the $N$–ion entangled state is obtained via quantum state reconstruction by expanding the density matrix in a basis of observables[16] and measuring the corresponding ex-pectation values. In order to do this, we employ additional laser pulses on the quadrupole transition to rotate the measurement basis prior to state detection[8]. We use $3^N$ different bases and repeat the experiment 100 times for each basis. For $N = 8$, this amounts to 656 100 experiments and a total measurement time of 10 hours. To obtain a positive semi–definite density matrix $\rho$, we follow the iterative procedure outlined by Hradil *et al.*[17] for performing a maximum–likelihood estimation of $\rho$. The reconstructed density matrix for $N = 8$ is displayed in Fig. 1. To retrieve the fidelity $F = \langle W_N | \rho | W_N \rangle$, we adjust the local phases such that $F$ is maximized. The local character of those transformations implies that the amount of the entanglement present in the system is not changed. We obtain fidelities $F_4 = 0.85, F_5 = 0.76, F_6 = 0.79, F_7 = 0.76$ and $F_8 = 0.72$ for the 4,5,6,7 and 8–ion W–states, respectively.

We investigate the influence of quantum projection noise on the reconstructed density matrix and quantities derived from it by means of a Monte Carlo simulation. Starting from the reconstructed density matrix, we simulate up to 100 test data sets taking into account the major experimental uncertainty, i.e. quantum projection noise. Then the test sets are analyzed and we can extract probability distributions for all observables from the resulting density matrices.

For an investigation of the entanglement properties, we associate each particle $k$ of a state $\rho$ with a (possibly spatially separated) party $A_k$. We shall be interested in different aspects of entanglement between parties $A_k$, i.e. the non–locality of the state $\rho$. A detailed entanglement analysis is achieved by investigating (i) the presence of genuine multipartite entanglement, (ii) the distillability of multipartite entanglement and (iii) entanglement in reduced states of two qubits.

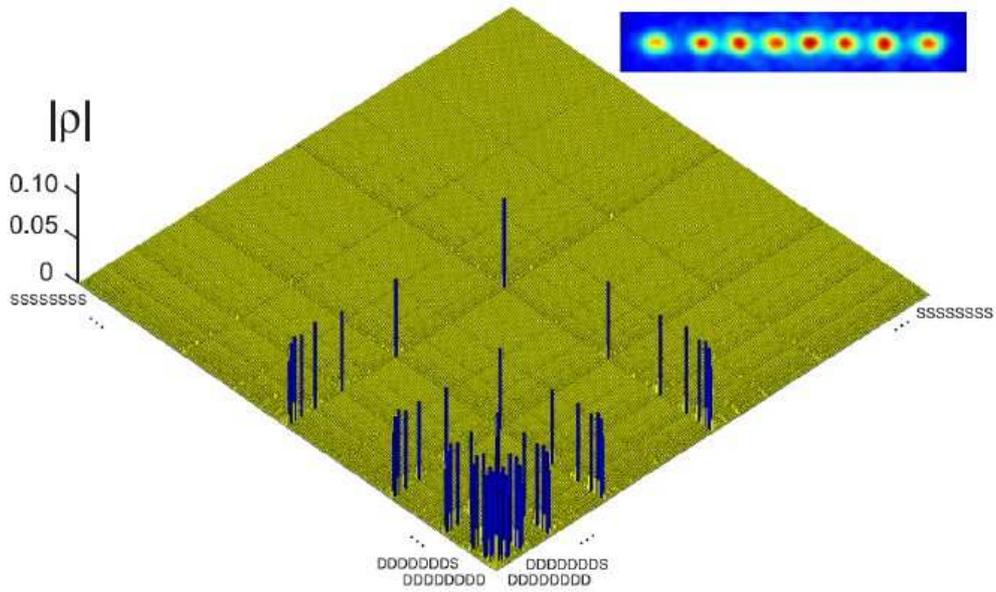

Figure 1: Absolute values of the reconstructed density matrix of a $|W_8\rangle$–state as obtained from quantum state tomography. Ideally, the blue coloured entries have all the same height of 0.125, the yellow coloured bars indicate noise. Numerical values of the density matrices for $4 \leq N \leq 8$ can be found in the supplementary information.

Firstly, we consider whether the production of a single copy of the state requires nonlocal interactions of all parties. This leads to the notion of genuine multipartite entanglement and biseparability. A pure multipartite state $|\psi\rangle$ is called biseparable, if two groups $G_1$ and $G_2$ comprised of parties $A_k$ can be found such that $|\psi\rangle$ is a product state with respect to the partition

$$|\psi\rangle = |\chi\rangle_{G_1} \otimes |\eta\rangle_{G_2}, \qquad (2)$$

otherwise it is genuinely multipartite entangled. A mixed state $\rho$ is called biseparable, if it can be produced by mixing pure biseparable states $|\psi_i^{bs}\rangle$ —which may be biseparable with respect to different bipartitions— with some probabilities $p_i$, i.e. the state can be written as $\rho = \sum_i p_i |\psi_i^{bs}\rangle\langle\psi_i^{bs}|$. If this is not the case, $\rho$ is genuinely multipartite entangled. In order to show the presence of multipartite entanglement, we use the method of entanglement witnesses [18–20]. An entanglement witness for multipartite entanglement is an observable with a positive expectation value on all biseparable states. Thus a negative expectation value proves the presence of genuine multipartite entanglement. A typical witness for the states $|W_N\rangle$ would be [20]

$$\mathcal{W}_N = \frac{N-1}{N} - |W_N\rangle\langle W_N|. \qquad (3)$$

This witness detects a state as entangled if the fidelity of the W–state exceeds $(N-1)/N$. However, more sophisticated witnesses can be constructed, if there is more information available on the state under investigation than only the fidelity. To do so, we add other operators to the witness in Eq. 3 (see Methods) which take into account that certain biseparable states can be excluded on grounds of the measured density matrix. Table 2 lists the expectation values for these advanced witnesses. The negative expectation values prove that in our experiment genuine four, five, six, seven and eight qubit entanglement has been produced.

Secondly, we consider the question whether one can use many copies of the state $\rho$ to distill one pure multipartite entangled state $|\psi\rangle$ by local means, i.e. whether entanglement contained in $\rho$ is qualitatively equivalent to multiparty pure state entanglement. For this aim one determines whether there exists a number $M$ such that the transformation

$$\underbrace{\rho \otimes \rho \otimes ... \otimes \rho}_{\text{M copies}} \stackrel{LOCC}{\longrightarrow} |\psi\rangle \qquad (4)$$

is possible. Here, $|\psi\rangle$ is a genuine multipartite entangled pure state (e.g. $|\psi\rangle=|W_N\rangle$) and $LOCC$ denotes a transformation using only local operations (with respect to the parties $A_k$) and classical communication. If such a transformation is possible, we call the state $\rho$ multipartite distillable [21]. Technically, multipartite distillability follows from the possibility to generate maximally entangled singlet states $|\psi^-\rangle = (|DS\rangle - |SD\rangle)/\sqrt{2}$ between any pair of parties $A_k, A_l$ by local means [21]. The latter can be readily shown for all reconstructed density matrices. Performing measurements of $\sigma_z$ on all particles except $k,l$ and restricting to outcomes $P_0 = |D\rangle\langle D|$ in all cases results in the creation of a two–qubit state $\rho_{kl}$. The density operator $\rho_{kl}$ is distillable entangled if the concurrence $C$, a measure for two–qubit entanglement [22], is non–zero. This is the case for all $k,l$ (see Tab. 2), which implies that $\rho_N$ is multiparty distillable entangled.

Thirdly, we investigate bipartite aspects of multiparticle entanglement [23], in particular the entanglement in the reduced states of two qubits. For W–states this is of special interest, since for these states all reduced density operators of two particles are entangled, and the entanglement is in fact maximal [4,24,25]. We investigate the bipartite entanglement by tracing out all but particles $k,l$ and obtain the reduced density operators $\rho'_{kl}$. From these density matrices we can now calculate the concurrence $C'_{kl} = C(\rho'_{kl})$ as a measure for the entanglement. For all $N$, we find that all reduced density operators are entangled (see Tab. 2). Note that the previous results (presence of multipartite entanglement and distillability) also imply that $\rho$ is inseparable and in fact distillable with respect to any bipartition $G_1 - G_2$ for all $N$.

Finally, we address the scalability of our approach. Four major sources for deviations from the ideal W–states are found: addressing errors, imperfect optical pumping, non–resonant excitations and frequency stability of the qubit–manipulation–laser (see Methods). All of them are purely technical and thus represent no fundamental obstacle for increasing the number of particles. Also the required blue sideband pulse area for a $|W\rangle$–state scales only with $\log N$ (see Tab. 1) while the time for a pulse with given area is proportional to the square root of the ion crystal's mass. Thus the overall favorable scaling behaviour of $\sqrt{N} \log N$ opens a way to study large scale entanglement experimentally.

**Methods**

Witnesses for our experiment can be derived as follows: For $N$ qubits we define the $N$ states $|BS_i\rangle = |D\rangle_i \otimes |W_{N-1}\rangle$, which consist of $|D\rangle$ on the $i$-th qubit and the state $|W_{N-1}\rangle$ on the remaining qubits. For the operator

$$\mathcal{Q} = \alpha |W_N\rangle\langle W_N| - \beta \sum_{i=1}^{N} |BS_i\rangle\langle BS_i|. \tag{5}$$

we then compute the maximal expectation value for biseparable states. Since mixed biseparable states are convex combinations of pure biseparable states, it suffices to look at pure biseparable states, thus we have to compute $\gamma = \max_{|\psi\rangle=|a\rangle\otimes|b\rangle}\langle\psi|\mathcal{Q}|\psi\rangle$ for all possible bipartitions[20]. If we investigate a partition where $|a\rangle$ is a $K$ qubit state, it can be seen that the optimal $|a\rangle$ is of the form $|a\rangle = a_0|DD\cdots D\rangle + b_1|DD\cdots DS\rangle + b_2|D\cdots DSD\rangle + \cdots + b_K|SDD\cdots D\rangle$. Then, from the matrix representation of $\mathcal{Q}$ one can deduce that the $a_0, b_1, \cdots, b_K$ can be chosen real and finally that $b_i = b_j$ for all $i, j$. A similar statement can be proven for $|b\rangle$, thus for an arbitrary number of qubits the optimization procedure can be reduced to a four parameter maximization with two normalization constraints, which can be efficiently solved numerically. The witness is then given by

$$\tilde{\mathcal{W}}_N = \gamma \mathbb{1}_2 - \mathcal{Q} \qquad (6)$$

where $\mathbb{1}_2$ denotes the identity operator on the space spanned by the elements of the computational basis which consists of $|D\rangle$ on at most two qubits. The adding of the term $\gamma\mathbb{1}_2$ guarantees that $\tilde{\mathcal{W}}_N$ is positive on all biseparable states. For the entanglement detection, we used the values $\alpha = 10$ and then $\beta = 2.98$, $\gamma = 2.2598$, for three qubits, $\beta = 2.87$, $\gamma = 0.8316$ for four qubits, $\beta = 2.35$, $\gamma = 0.3760$ for five qubits, $\beta = 1.94$, $\gamma = 0.1937$ for six qubits, $\beta = 1.638$, $\gamma = 0.1139$ for seven qubits, $\beta = 1.4125$, $\gamma = 0.0764$ for eight qubits, .

For $N = 8$ we have in addition optimized the witness using local filtering operations, i.e., we applied a transformation $\tilde{\mathcal{W}}_8^f = F\tilde{\mathcal{W}}_8 F^\dagger$ with $F = F_1 \otimes F_2 \otimes \cdots \otimes F_8$. Here the $F_i$ are operators acting on each qubit separately and are thus local operations. Therefore the new witness $\tilde{\mathcal{W}}_8^f$ remains positive on all biseparable states. Finally, all witnesses have been normalized such that their expectation value for the maximally mixed state equals one and the local phases have been adjusted.

For an investigation of the experimental imperfections and scalability, we simulate the preparation procedure by solving the Schrödinger equation with all relevant imperfections. This way we identify four major sources of deviations from the ideal W–states: addressing errors, imperfect optical pumping, non–resonant excitations, and laser frequency noise. The trap frequency influences these experimental imperfections diametrically: for example, to keep the addressing error reasonably low [i.e. less than 5%, where the addressing error is defined as the ratio of the Rabi–frequencies between the addressed ion and the neighboring ion(s)], we adjust the trap frequency such that the inter–ion distance in the center of the ion string is about 5 $\mu$m. However, for large $N$ the required trap relaxation implies that the sideband transition frequency moves closer to the carrier transition frequency. Thus the strong laser pulses driving the weak sideband transition cause more off–resonant excitations on the carrier transition, which in turn spoil the obtainable fidelity. Therefore we reduce the laser power for driving the sideband, which then results in longer preparation times and leads to an enhanced susceptibility to laser frequency noise. A compromise for the different ion numbers $N$ is the following set of parameters: ($N = 4$: $\nu = 1.123$ MHz, $T_{2\pi} = 220\,\mu$s), ($N = 5$: $\nu = 1.055$ MHz, $T_{2\pi} = 300\,\mu$s), ($N = 6$: $\nu = 0.905$ MHz, $T_{2\pi} = 350\,\mu$s), ($N = 7, 8$: $\nu = 0.813$ MHz, $T_{2\pi} = 380\,\mu$s). Here $\nu$ is the trap frequency (center of mass) and $T_{2\pi}$ is the time for a $2\pi$–pulse on the blue sideband. The fidelity reduction of $|W_6\rangle$ for the different imperfections are as follows: 0.1 (addressing error), 0.07 (off–resonant excitations), 0.04 [laser frequency noise (200 Hz rms)]. Another possible error source is imperfect ground state cooling. Intensity noise of the 729–laser ($\Delta I_{\max}/I \approx 0.03$) does not contribute significantly. Finally, we experimentally observed non–ideal optical pumping which can result in a reduction of 0.02 of the fidelity per ion. For $N \geq 6$, we therefore minimize the errors due to optical pumping and a part of the addressing errors by checking the initialization procedure with a detection sequence (see Tab. 1).

**Supplementary Information** Reconstructed density matrices for $|W_N\rangle$ ($4 \leq N8 \leq N$), witness operator for $N = 8$. Raw data is available upon request.

**Acknowledgments** We gratefully acknowledge support by the Austrian Science Fund (FWF), by the European Commission (QGATES, CONQUEST PROSECCO, QUPRODIS and OLAQUI networks), by the Institut für Quanteninformation GmbH, the DFG, and the ÖAW through project APART (W.D.). This material is based upon work supported in part by the U. S. Army Research Office. We thank P. Pham for the pulse modulation programmer and A. Ostermann, M. Thalhammer and M. Ježek for the help with the iterative reconstruction.

**Competing Interests** The authors declare that they have no competing financial interests.

**Correspondence** Correspondence and requests for materials should be addressed to H.H. (email: hartmut.haeffner@uibk.ac.at).

(i1) $\left| \begin{array}{l} |0, SSS \cdots S\rangle \\ \xrightarrow{R_N^C(\pi) R_{N-1}^C(\pi) \cdots R_1^C(\pi)} \\ |0, DDD \cdots D\rangle \\ \text{Check state via fluorescence} \end{array} \right.$

(i2) $\left| \begin{array}{l} \xrightarrow{R_1^+(\pi)} \\ |0, DDD \cdots D\rangle \\ \text{Check state via fluorescence} \end{array} \right.$

(i3) $\left| \begin{array}{l} \xrightarrow{R_N^C(\pi)} \\ \frac{1}{\sqrt{N}} |0, SDD \cdots D\rangle \end{array} \right.$

(1) $\left| \begin{array}{l} \xrightarrow{R_N^+(2 \arccos(1/\sqrt{N}))} \\ \frac{1}{\sqrt{N}} |0, SDD \cdots D\rangle + \frac{\sqrt{N-1}}{\sqrt{N}} |1, DDD \cdots D\rangle \end{array} \right.$

(2) $\left| \begin{array}{l} \xrightarrow{R_{N-1}^+(2 \arcsin(1/\sqrt{N-1}))} \\ \frac{1}{\sqrt{N}} |0, SDD \cdots D\rangle + \frac{1}{\sqrt{N}} |0, DSD \cdots D\rangle + \frac{\sqrt{N-2}}{\sqrt{N}} |1, DDD \cdots D\rangle \end{array} \right.$

⋮

$\left| \begin{array}{l} \frac{1}{\sqrt{N}} |0, SDD \cdots D\rangle + \frac{1}{\sqrt{N}} |0, DSD \cdots D\rangle + \cdots + \frac{1}{\sqrt{N}} |1, DDD \cdots D\rangle \end{array} \right.$

(N) $\left| \begin{array}{l} \xrightarrow{R_1^+(2 \arcsin(1/\sqrt{1}))} \\ \frac{1}{\sqrt{N}} |0, SDD \cdots D\rangle + \frac{1}{\sqrt{N}} |0, DSD \cdots D\rangle + \cdots + \frac{1}{\sqrt{N}} |0, DDD \cdots S\rangle \end{array} \right.$

Table 1: Creation of a $|W_N\rangle$–state ($N = \{6, 7, 8\}$). First we initialize the ions via sideband cooling and optical pumping in the $|0, SS \cdots S\rangle$–state where we use the notation $|n, x_N x_{N-1} \cdots x_1\rangle$. $n$ describes the vibrational quantum number of the ion motion and $x_i$ their electronic state. We then prepare the $|0, DDD \cdots D\rangle$–state with $N$ $\pi$–pulses on the carrier transition applied to ions #1 to #$N$, denoted by $R_n^C(\theta = \pi)$ (the notation is detailed in Gulde *et al.*[15]; we do not specify the phase of the pulses because its particular value is irrelevant in this context). Then this state is checked for vanishing fluorescence with a photomultiplier tube. The same is done after trying to drive a $\pi$–pulse on the blue sideband on ion #1 to ensure that the ion crystal is in the motional ground state. After this initialisation, we transform the state to $|0, SDD \cdots D\rangle$ with a carrier pulse and start the entanglement procedure in step (1). This is carried out by moving most of the population to the $|1, DDD \cdots D\rangle$ with a blue sideband pulse of length $\theta_n = \arccos(1/\sqrt{n})$ leaving the desired part back in $|0, SDD \cdots D\rangle$. Finally, we use $N - 1$ blue sideband pulses ($R_n^+(\theta_n)$) of pulse length $\theta_n = \arcsin(1/\sqrt{n})$ such that at each step we split off a certain fraction of the wave packet. Note that for an ion string in the ground state, blue–sideband pulses acting on an ion in the D–state have no effect. For $N = \{4, 5\}$ we do not check the fluorescence, combine steps $i1$ and $i3$ and omit step $i2$.

|              | $N=3$  | $N=4$       | $N=5$       | $N=6$       | $N=7$       | $N=8$       |
|:------------:|:------:|:-----------:|:-----------:|:-----------:|:-----------:|:-----------:|
| $F$ | 0.824 | 0.846 (11) | 0.759 (7) | 0.788(5) | 0.763 (3) | 0.722 (1) |
| $\mathrm{tr}(\tilde{\mathcal{W}}_N \rho_N)$ | $-0.532$ | $-0.460\,(31)$ | $-0.202\,(27)$ | $-0.271\,(31)$ | $-0.071\,(32)$ | $-0.029\,(8)$ |
| $\min(C_{kl})$ | 0.724 | 0.760 (34) | 0.605 (23) | 0.567 (16) | 0.589 (9) | 0.536 (8) |
| $\bar{C}$ | 0.776 | 0.794 (23) | 0.683 (15) | 0.677 (11) | 0.668 (5) | 0.633 (3) |
| $\min(C'_{kl})$ | 0.294 | 0.229 (21) | 0.067 (12) | 0.049 (4) | 0.035 (4) | 0.022 (3) |
| $\bar{C}'$ | 0.366 | 0.267 (12) | 0.162 (6) | 0.124(3) | 0.091 (2) | 0.073 (1) |

Table 2: Entanglement properties of $\rho_N$. First row: Fidelity after properly adjusting local phases. Second row: Expectation value of the witnesses $\tilde{\mathcal{W}}_N$ (for $N=8$ we used additionally local filters). Third and fourth row: minimal and average concurrence between two qubits after $\sigma_z$–measurement on the remaining $(N-2)$ qubits. Fifth and sixth row: minimal and average concurrence between two qubits after discarding the remaining $(N-2)$ qubits. For completeness we also analyzed the data published previously in Ref. [6] for $N=3$.